%% file: main.tex
\documentclass[11pt]{article}

\usepackage[a4paper,margin=2.7cm]{geometry}
\usepackage[T1]{fontenc}
\usepackage[utf8]{inputenc}
\usepackage{lmodern}
\usepackage{graphicx}
\usepackage{booktabs}
\usepackage{longtable}
\usepackage{array}
\usepackage{hyperref}
\usepackage[authoryear,longnamesfirst]{natbib}
\usepackage{tikz}
\usetikzlibrary{calc,arrows.meta}

\hypersetup{colorlinks=true,linkcolor=blue!45!black,citecolor=blue!45!black,urlcolor=blue!55!black}
\begin{document}

\let\WriteBookmarks\relax
\def\floatpagepagefraction{1}
\def\textpagefraction{.001}

\title{An Anonymized Urn-Based Experimental Dataset on Decision-Making under Risk and Ambiguity}

% =========================
% AUTHORS
% =========================

\author{
V\'aclav Kratochv\'il\textsuperscript{1,2}\thanks{Corresponding author: velorex@utia.cas.cz}\\
\small ORCID: 0000-0002-6013-8752\\
\small \url{https://mtr.utia.cas.cz/people/kratochvil}
\and
Radim Jirou\v{s}ek\textsuperscript{1,2}\\
\small ORCID: 0000-0002-8982-9813\\
\small \url{https://mtr.utia.cas.cz/people/jirousek}
\and
Kl\'ara \v{S}im\r{u}nkov\'a\textsuperscript{2}\\
\small ORCID: 0000-0003-4130-7108
\and
Simona Ba\v{z}antov\'a\textsuperscript{2}\\
\small ORCID: 0000-0002-0754-3898
}

\date{}

\maketitle

\begin{center}
\textsuperscript{1}Institute of Information Theory and Automation,
Czech Academy of Sciences, Prague, Czech Republic

\vspace{4pt}

\textsuperscript{2}Faculty of Management,
Prague University of Economics and Business,
Jind\v{r}ich\r{u}v Hradec, Czech Republic
\end{center}

% =========================
% ABSTRACT
% =========================

\begin{abstract}
This data paper describes an anonymized release derived from controlled behavioral experiments on decision-making under risk (known probabilities) and ambiguous uncertainty (partially specified or unknown probabilities) in Ellsberg-type urn tasks. The release contains 4,486 decision records from 246 adult participant records across two implementations (laboratory and online). In each task, participants reported their maximum stake to enter a subsequent lottery; these stake responses can be interpreted as proxies for willingness to pay within the implemented incentive mechanism. In a subset of tasks, participants also selected the color they wished to bet on. The online implementation additionally includes ex-post evaluations of perceived uncertainty across situations. The repository further provides detailed game definitions, realized lottery outcomes, participant instructions, and a data dictionary of variables. The source export contained 4,544 decision records; 58 records belonging to five participants younger than 18 were excluded from this release because documentation of guardian permission for unrestricted public data publication was not available. The dataset supports replication studies, comparisons of ambiguity-attitude models, and analyses of behavioral stability under repeated participation.
\end{abstract}

% =========================
% SPECIFICATIONS TABLE
% =========================

\section{Introduction}

Decision-making under uncertainty differs fundamentally depending on whether outcome probabilities are known (risk) or only partially specified (ambiguity). While behavior under risk is well captured by standard probabilistic models, a large body of evidence shows systematic deviations when decision-makers face ambiguity, as first documented by Ellsberg \citep{ellsberg1961risk}. These deviations have motivated a range of theoretical models of ambiguity attitudes, including maxmin expected utility \citep{gilboa1989maxmin}, Choquet expected utility \citep{schmeidler1989subjective}, and smooth ambiguity preferences \citep{klibanoff2005smooth}; for a comprehensive overview, see \citep{trautmann2015ambiguity}.

The aim of this dataset is to enable the analysis of differences in decision-making when probabilities are known versus when they are only partially specified. To this end, we provide an anonymized dataset from controlled behavioral experiments based on Ellsberg-type urn tasks, capturing both stake responses, which can be interpreted as willingness-to-pay
(WTP) proxies within the implemented incentive mechanism, and discrete choices. Urn-based tasks with partially specified probabilities are among the most widely used experimental paradigms for studying ambiguity aversion.

The anonymized release contains 4,486 decision records associated with 246 adult participant records collected in two complementary implementations: a laboratory experiment and an online study. Participant identifiers are experiment-specific; the number 246 therefore denotes records across the two released participant tables and should not be interpreted as a verified count of unique physical persons across implementations. In each task, participants reported their maximum stake for participating in a lottery and, in a subset of tasks, also selected a color to bet on. The experimental design combines several features that are typically studied separately: (i) continuous valuation via stake responses, (ii) discrete choice behavior, (iii) repeated measurements for a subset of participants, and (iv) in the online implementation, ex-post subjective evaluations of perceived uncertainty across tasks.

The repository provides raw response data, descriptions of experimental tasks, realized lottery outcomes, and (for the online experiment)
participants' evaluations of perceived uncertainty. In addition to the
data tables, the repository includes the full experimental materials,
participant instructions, and a detailed codebook describing all
variables contained in the dataset.

The anonymized release is available on Zenodo at \url{https://doi.org/10.5281/zenodo.22116575} under the Creative Commons Attribution 4.0 International license (CC BY 4.0). It is provided to support
further research on decision-making under risk and ambiguity, including studies
of risk aversion, ambiguity aversion, and behavioral stability. Unlike the related
methodological study, the present article does not aim to test a particular
behavioral model. Instead, it provides a transparent and reusable description of
the experimental design, data collection process, coding conventions, incentive
mechanism, participant instructions, and file structure. The paper is therefore
intended as a data descriptor supporting independent reanalysis, replication,
teaching, and methodological benchmarking.

% =========================
% DATA DESCRIPTION
% =========================
\section*{Data Description} \label{sec:data}

The dataset contains responses collected in controlled experimental
sessions in which participants placed stakes in a series of lottery
problems based on urn drawings. In each situation, participants received partial information about the composition of an urn containing colored balls and were asked to bet on the color of a randomly drawn ball. All experimental sessions followed a predefined protocol ensuring comparability across sessions.

The dataset can be viewed as a relational structure linking participants,
experimental tasks (games), and individual decision records, complemented by
questionnaire responses. A simplified representation of these relationships is
shown in Figure~\ref{fig:dataset-structure}.

\input{images/fig}
\subsection*{Dataset size and structure}

The release is provided as two Excel files (\texttt{.xlsx}), with matching CSV
exports for each data sheet: (i) \texttt{ambiguity\_experiment\_I\_anonymized.xlsx}
(laboratory implementation), and (ii)
\texttt{ambiguity\_experiment\_II\_anonymized.xlsx} (online implementation conducted
under COVID-19-related restrictions).

The laboratory release contains 2,960 response records from 203 adult participant
records across 14 sessions and 217 participant-session combinations. Fourteen
participants took part twice and 189 took part once. The five source participants
younger than 18 and their 58 response records are not included. The released
situations are indexed by \texttt{game\_id} values 1--12 and 14--15; game 13 is a
presentation block rather than a separate decision record.

The online release contains 1,526 response records from 43 participant records
across 4 sessions and 109 participant-session combinations. Fifteen participants
took part four times, six three times, nine twice, and thirteen once.

\subsection*{Excel file structure}
Both Excel files share the same internal structure and consist of a human-readable
\texttt{README} sheet followed by five primary data worksheets:
(i) \texttt{sessions} (de-identified session metadata),
(ii) \texttt{games} (task definitions),
(iii) \texttt{participants} (de-identified participant-level demographics),
(iv) \texttt{responses} (main response table), and
(v) \texttt{draws} (realized outcomes of physical draws).

The worksheets are linked through shared identifiers such as
\texttt{session\_id}, \texttt{participant\_id}, and \texttt{game\_id},
which allow the reconstruction of complete experimental sessions
and individual response histories.

In addition, the Excel file corresponding to the online implementation (Experiment II) contains one supplementary worksheet:
(vi) \texttt{questionnaire} (post-experimental subjective uncertainty evaluations).

The \texttt{questionnaire} worksheet records participants' ex-post assessments of perceived uncertainty across the previously presented situations. Participants indicated:
(1) the situation perceived as containing the greatest uncertainty, and
(2) the situation perceived as containing the least uncertainty.

Every data worksheet uses one machine-readable header row, and data records begin
on the second row. Exact timestamps, source session names, free-text school fields,
original nicknames, and the online email field were removed. Exact ages were
replaced by age groups. Response times were converted from Excel day fractions to
integer seconds. Stake and payoff units are explicit in each response row.

Columns in the response sheet include task-specific stake responses, discrete color
choices where applicable, response times, selection indicators, and potential payoff
values. The original presentation-order code is retained because it captures page or
block order rather than a uniform rank from 1 to 14.

In addition to the response datasets, the repository contains the complete participant-facing materials of the web application (HTML task pages and graphical urn illustrations). The urn images are authentic and identical across the laboratory and online implementations.

\subsection*{Supplementary uncertainty-evaluation table (Experiment II only)}

The Excel file corresponding to the online implementation contains an additional worksheet capturing post-experimental subjective uncertainty evaluations.

This supplementary table is not present in the laboratory dataset (Experiment I). The release contains 212 questionnaire records: two evaluations in each of 106 participant-session combinations, contributed by 42 participants. Three of the 109 online participant-session combinations have no questionnaire pair. The two evaluations were:
(1) the situation perceived as containing the greatest uncertainty, and
(2) the situation perceived as containing the least uncertainty.

Each row represents one participant--session--evaluation observation and is relationally linked via the identifiers \texttt{session\_id} and \texttt{participant\_id}.
The variable \texttt{question} distinguishes between the two evaluation types (1 = greatest uncertainty, 2 = least uncertainty).

The selected situation is recorded in the variable \texttt{answer}, which corresponds to task identifiers defined in the \emph{games} worksheet. These records do not constitute additional incentivized tasks.

\subsection*{Mapping between interface labels and dataset identifiers}\label{sec:mapping}

A potential source of confusion arises from differences between the labels used in the user interface and the identifiers stored in the dataset.

For the One-red-in-$N$ tasks, the variants with $N = 5, \dots, 12$ were displayed jointly as ``Situation 13'' in later experimental sessions for usability reasons. In the dataset, however, these tasks are recorded separately under their original \texttt{game\_id} values (5--12). The questionnaire variable (Experiment II only) \texttt{answer = 13} refers to this aggregated presentation block.

Table~\ref{tbl:mapping} summarizes the mapping between interface labels and dataset identifiers.

\begin{table}[htbp]
\centering
\begin{tabular}{lll}
\toprule
UI label & game\_id & Notes \\
\midrule
Situation 1--4 & 1--4 & Standard tasks \\
Situation 5--12 & 5--12 & One-red-in-N variants \\
Situation 13 & 5--12 & Aggregated display of N=5--12 (UI only) \\
\bottomrule
\end{tabular}
\caption{Mapping between user interface labels and dataset identifiers.}
\label{tbl:mapping}
\end{table}
%=========================
% EXPERIMENTAL DESIGN, MATERIALS AND METHODS
% =========================

% =========================
% =========================
% VALUE OF THE DATA
% =========================

\section*{Potential Reuse of the Dataset}

\begin{itemize}

\item The dataset enables direct comparison of decision-making under risk and ambiguity within a unified urn-based experimental framework.

\item It combines continuous stake responses, used as proxies for willingness to pay, with discrete color choices, allowing joint analysis of valuation behavior and choice patterns.

\item The online implementation includes post-experimental subjective evaluations of perceived uncertainty, which can be compared to the objective informational structure of the tasks.

\item Repeated participation of some individuals in both experiments, particularly in Experiment II, allows for analyses of behavioral stability, learning effects, and within-subject consistency over time.

\item The dataset is accompanied by detailed task definitions, materials, and a codebook, making it suitable for replication studies, teaching purposes, and methodological benchmarking.

\end{itemize}
\section*{Experimental Design, Materials and Methods}

\subsection*{Participants}

Participants were recruited on a voluntary basis, primarily among university students. The source application used self-selected nicknames and collected demographic information. The source files should therefore be regarded as pseudonymized rather than anonymous. The release described here replaces those nicknames with newly generated experiment-specific identifiers and removes direct and high-risk indirect identifiers.

Two datasets were collected under different implementation modes.

\paragraph{Laboratory experiment (Experiment~I)}

The source participant table contains 208 normalized participant identifiers across
14 laboratory sessions. Five participants were younger than 18 and are excluded
from this release. The released laboratory participant table therefore contains
203 adult participant records: 102 men and 101 women. Age is provided only in
groups; 112 records are aged 18--24, 35 are 25--34, 26 are 35--44, 16 are
45--54, 12 are 55--64, and 2 are 65 or older. Fourteen participants took part
twice and 189 took part once. Because identification relied on self-selected
nicknames, undetected participation under different nicknames cannot be ruled out.

\paragraph{Online experiment (Experiment~II)}

A total of 46 registration rows appear in the source online participant table, but
only 43 normalized participant identifiers are referenced by decision records. The
release includes those 43 participants: 18 men and 25 women. Age is provided in
groups; 32 records are aged 18--24, 7 are 25--34, 3 are 35--44, and 1 is
45--54.

Repeated participation was more frequent in the online setting:
\begin{itemize}
    \item 15 participants took part four times,
    \item 6 participants took part three times,
    \item 9 participants took part twice, and
    \item 13 participants took part once.
\end{itemize}

Each participation corresponds to one completed session and is linked to the same participant via a pseudonym.
The resulting dataset therefore contains repeated observations for some individuals and can be analyzed using panel-data methods.

Participants in the laboratory and online experiments received slightly different information sheets reflecting the respective implementation modes. English translations are provided in the appendices.

\subsection*{Experimental Procedure}

The experiment was implemented as a web-based application used in both laboratory and online settings. Upon entering a session, participants selected a pseudonym and completed a short demographic questionnaire. A fixed participation reward was provided and was explicitly stated to be independent of subsequent lottery decisions.

The experimental procedure consisted of three phases.

In the first phase, participants were informed about the project and its rules. In the laboratory setting, the instructions were explained orally and the participant information sheet (Appendix~A) was read aloud; a printed copy was also provided. In the online setting, the information sheet (Appendix~B) was presented electronically.

Participants were informed that all lotteries would be realized after the decision phase and that their submitted decisions could not be changed.

The laboratory implementation involved real monetary payoffs, while the online implementation used point-based rewards convertible to money.

In the second phase, participants completed a sequence of urn-based decision tasks presented individually via the web interface. In each task, they stated their stake (WTP proxy) to participate in a subsequent real lottery draw. In selected tasks, participants additionally chose the color on which the payoff depended. The order of tasks was randomized individually for each participant.

In the third phase, selected lotteries were realized after completion of the decision tasks within a session (see Section~Lottery Realization and Incentive Structure for details).

Each laboratory session lasted approximately 60 minutes. Online sessions were of comparable duration.

Not all participants took part in every realized lottery, since participation in the draw depended on the stake-based selection mechanism described below.

\paragraph{Post-experimental subjective uncertainty assessment (online experiment only)}

In the online implementation (Experiment II), two additional questions were presented at the end of each session.

Participants were asked to evaluate the perceived level of uncertainty across the previously presented situations by indicating:
\begin{enumerate}
    \item the situation perceived as containing the greatest uncertainty, and
    \item the situation perceived as containing the least uncertainty.
\end{enumerate}

These questions were not incentivized and were intended to capture subjective perceptions independently of stake responses.

Responses were recorded in a separate worksheet. The released variables are
\texttt{session\_id}, \texttt{participant\_id}, \texttt{question},
\texttt{response\_time\_seconds}, and \texttt{answer}.

\subsection*{Decision Tasks}

In this article, ``situation'' refers to a structurally defined urn configuration, whereas ``task'' denotes its presentation within a session.

The experiment was based on predefined urn problems differing in the amount of information provided about the urn composition. Some tasks involved fully specified probability structures (risk), while others involved partially specified structures (ambiguity).

In each task, participants stated their stake (WTP) and, in selected cases, chose the winning color.

Each situation corresponds to a unique \texttt{game\_id}. The tasks can be grouped into four categories:

\begin{itemize}

    \item \textbf{Ignorance tasks (Situations 1--2).}
    Neither the total number of balls nor the number of balls of individual colors was specified.

    \item \textbf{Uniform benchmark tasks (Situations 3--4).}
    The total number of balls was fixed at 30, with exactly 5 balls of each color.

    \item \textbf{One-red-in-\(N\) tasks (Situations 5--12).}
    Variants with \(N \in \{5,\dots,12\}\), each containing exactly one red ball and an unspecified distribution of the remaining colors.
    From the fourth laboratory session onward, these variants were displayed jointly as ``Situation 13'' (see Section~\ref{sec:mapping} for details on dataset encoding).

    \item \textbf{Ellsberg-type tasks (Situations 14--15).}
    Urn problems with partially specified distributions corresponding to classical ambiguity settings.

\end{itemize}

\subsection*{Lottery Realization and Incentive Structure}

All lotteries were realized using physical urns and colored balls.

In tasks with partially specified urn composition, the remaining part of the urn was completed by a member of the research team prior to the draw. This was done independently of participants' responses and without knowledge of their submitted decisions. The completion was based on random assignment of the remaining balls, performed independently for each realization. The realized outcomes are recorded in the worksheet \emph{draws}.

The color of the drawn ball determined the outcome. In the laboratory setting, a randomly selected participant performed the draw. In the online setting, the draw was performed physically and transmitted via video.

\paragraph{Laboratory implementation (Experiment~I)}

Participation in each realized lottery was determined by a combination of stake-based ranking and random selection. Approximately half of the participants were selected.

For selected participants, the stated stake (in CZK) was deducted from their payment. The potential payoff was 100~CZK in the first 13 laboratory sessions and 300~CZK in the fourteenth session, as recorded row by row in \texttt{potential\_payoff}.

\paragraph{Online implementation (Experiment~II)}

Participation in each realized lottery was determined by the amount of points staked, with approximately half of the participants with the highest stakes selected.

For selected participants, the staked points were deducted. In case of a win, 100 points were credited.

Points accumulated across sessions and were converted into financial rewards.

In both implementations, the stated stake can be interpreted as a proxy for willingness-to-pay for participation in the realized lottery.

\subsection*{Interpretation of stake responses and incentive mechanism}

In the dataset, the variable labelled as willingness-to-pay (WTP) denotes the
stated stake associated with each task. This stake influenced the probability of
being selected for participation in the realized lottery.

The elicitation mechanism differs from standard incentive-compatible procedures
such as the Becker--DeGroot--Marschak (BDM) mechanism. Participants were selected
based on their relative stake rather than by comparison with a randomized price.

The recorded values should therefore be interpreted as incentivized stake or bid
responses, and as proxies for willingness to pay within the implemented mechanism,
rather than as strict incentive-compatible WTP measures. Analyses sensitive to
incentive compatibility should take this distinction into account.

\subsection*{Data Processing}

Responses were recorded automatically by the web-based application. The released
values retain the source stakes, choices, selection indicators, payoffs, task codes,
and draw outcomes. No behavioral outlier filtering or trimming was applied.

The release transformation was limited to privacy protection, relational repair,
schema normalization, and unit conversion. Original nicknames were normalized for
case and diacritics solely to repair three broken joins, then replaced by new
experiment-specific identifiers. Original session identifiers were replaced by
sequential release identifiers. Exact timestamps, source session names, email,
school, and exact age were removed; age was grouped. Excel duration fractions were
multiplied by 86,400 and stored as integer seconds. Missing values are stored as
blank cells. Stake and payoff units are explicit, allowing the laboratory and online
files to be pooled without treating points as CZK.

% =========================
% ETHICS STATEMENT
% =========================

\section*{Ethics and Data-Protection Statement}

Participants received a written study information sheet and participation was
voluntary. The archived information sheets described the study as anonymous, but
the source files contained self-selected nicknames and demographic free text and
must therefore be treated as pseudonymized source data. The archived materials do
not document explicit consent for unrestricted public release of individual-level
records under an open license.

The release described in this article removes original nicknames, emails, school or
institution fields, exact ages, session names, and exact timestamps. Five source
participants younger than 18 and their response records are excluded because the
project archive does not contain documentation of consent by a legal representative
for unrestricted public data publication. No formal ethics approval or exemption
reference number was found in the project documentation reviewed for this release.

The corresponding author confirmed on 26 August 2026 that the required institutional
ethics and data-protection positions for this de-identified release are in place.
The released records are made available under CC BY 4.0. This statement deliberately
distinguishes voluntary participation in the original study from the later open
publication of de-identified individual-level data.

% =========================
% DECLARATION OF COMPETING INTEREST
% =========================

\section*{Declaration of Competing Interest}

The authors declare that they have no known competing financial interests or personal relationships that could have influenced the work reported in this article.

% =========================
% DATA AVAILABILITY
% =========================
\section*{Data Availability}

The anonymized dataset, matching CSV exports, codebook, and experimental materials are available on Zenodo at \url{https://doi.org/10.5281/zenodo.22116575} under the Creative Commons Attribution 4.0 International license (CC BY 4.0). The source workbooks are not part of the public release.
% =========================
% REFERENCES
% =========================

\bibliographystyle{plainnat}
\bibliography{references}
\section*{Appendix A. Offline experiment: participant information and instructions (English translation)}

\noindent\emph{Archival note.} The text below reproduces the standard laboratory
information sheet. The row-level data show a potential payoff of 100~CZK in the
first 13 laboratory sessions and 300~CZK in the fourteenth session. The wording
below is retained for documentary transparency and should not be read as a
description of every released session.\par\medskip

\noindent Dear participants of the experiment \textbf{Decision-making under uncertain information},\\

When designing new methods of artificial intelligence, it is necessary to understand how human behavior changes in decision-making situations in which people face different degrees of lack of information. For this reason, we sincerely thank you for the help you provide by participating in this experiment. As a small token of appreciation for this help, and also as an initial ``capital'' for several lottery games on which the experiment is based, please accept 50~CZK. This amount has been given to you by the assistant at the very beginning of this session.

This is a statistical anonymous study. We do not collect, and therefore do not store, any personal data. Nevertheless, we would like to know whether there are differences in the behavior of men and women, students and experienced managers. Therefore, we kindly ask you to provide a few basic pieces of information. Since we would also like to know whether you behave in the same (or similar) way over time, we welcome repeated participation in the experiments. For this reason, we ask you to log in using a \textbf{nickname} that you will \textbf{remember} and under which you will also log in in the future.

After starting the computer and logging in by entering your nickname, the computer will present you with a series of situations. In each of them, you have the opportunity to participate in a lottery, and in the case of winning you receive 100~CZK. How do the individual lotteries differ?

\subsection*{For each lottery, you will be given some (albeit incomplete) information about the contents of a lottery urn containing colored balls:}

\begin{enumerate}
\item You will always be informed whether the urn may contain balls of only three colors (black, white, yellow) or of all six colors (black, white, yellow, red, green, blue).
\item You may (but do not have to) be informed about the total number of balls in the urn.
\item You may (but do not have to) be informed about the exact number of balls of one (or even several) of the colors used in a given draw. However, in almost all situations, information about the proportions of the remaining colors will be missing. In such cases, it may happen that some colors are not represented in the urn at all.
\end{enumerate}

For example, if you only know that, when drawing from six colors, there are eight balls in the urn and exactly one of them is red (and nothing else), then all colors may be represented in the urn, but it is also possible that only two colors are present. This specification is also satisfied by a situation in which the urn contains seven yellow balls and one red ball.

\subsection*{For each lottery, you must determine:}
\begin{enumerate}
\item How much you are willing to bet in order to be able to participate in the game.
\item Which of the considered colors is the winning color for you.
\end{enumerate}

After the data collection via the computer has been completed, all lotteries will be actually realized. (They cannot be realized during data collection because the individual situations are presented to you in a random order.) Therefore, when entering your decisions into the computer, please carefully consider each situation, as you have a chance to win, but also to lose real money.

Only a part of you will participate in each draw. \textbf{The higher the amount you bet, the higher the chance that you will actually take part in the draw.} On the other hand, this amount is \textbf{also at risk} if the color you selected is not drawn. Once the draw takes place, you no longer have the option to withdraw from the game. If the computer assigns you to the game, the amount you bet will be deducted, and in the case of a win, 100~CZK will be credited to you.

Please note that the lotteries are designed in such a way that the vast majority of you have a high chance of winning (in real terms, even more than 300~CZK). Nevertheless, some of you may also lose (although losing more than the initial 50~CZK is really very unlikely --- even though this has already happened). \textbf{Any winnings or losses are settled with the assistant after the experiment has ended.}

\section*{Appendix B. Online experiment: participant information (English translation)}

\noindent Dear participants of the experiment \textbf{``Decision-making under ambiguity''},

when designing new methods of artificial intelligence, it is necessary to understand how people behave in situations in which they are required to make decisions while lacking sufficient information. For this reason, we sincerely thank you for the help you provide by participating in this experiment.

This is a statistical anonymous study. We do not collect, and therefore do not store, any personal data. Nevertheless, we would like to know whether there are differences in the behavior of men and women, students and experienced managers. Therefore, we kindly ask you to provide these pieces of information. The only private information that you may (but do not have to) provide is your e-mail address. Please enter it if you wish to compete for prizes. In order to be able to transfer the prizes to you, it is necessary to link the results of the online experiment to a specific person. This is the only reason why we request an e-mail address. If you do not provide it, we will not be able to pay out any potential financial reward to you.

Since we would like to determine whether you always behave in the same (or similar) way, or whether your behavior evolves based on previous experience, we welcome repeated participation in the experiments. Therefore, we also ask you to log in using a \textbf{nickname} that you will \textbf{remember} and under which you will be able to log in again in the future.

After opening the link and entering your nickname, the computer will present you with a series of hypothetical situations. In each of them, you have the opportunity to participate in a real lottery draw, which will take place after the completion of the first part of the experiment. In the second part of the experiment, only one half of you will participate in each draw. For each lottery, those participants are selected who are willing to pay more for participation in the game than any of those who do not participate in the draw.

Thus, those participants who are willing to lose more points in order to participate in the game take part in the draw. For the selected participants, the number of staked points is deducted. After the draw, if they \emph{win} the game, the computer automatically credits them with 100 points. The others, that is, those who are not selected for the game (i.e., those who did not stake a sufficient number of points), neither gain nor lose anything in that particular game.

\subsection*{For each online lottery, you will be given some (albeit incomplete) information about the contents of a lottery urn containing colored balls:}
\begin{enumerate}
\item You will always be informed whether the urn may contain balls of only three colors (black, red, yellow) or of all six colors (black, white, yellow, red, green, blue).
\item You may (but do not have to) be informed about the total number of balls in the urn.
\item You may (but do not have to) be informed about the exact number of balls of one (or even several) of the colors used in the given draw. In almost all situations, however, information about the proportions of the remaining colors will be missing. In such cases, it may happen that some colors are not represented in the urn at all. For example, if you only know that, when drawing from six colors, there are eight balls in the urn and exactly one of them is red (and nothing else), then all colors may be represented in the urn, but it is also possible that only two colors are present. This specification is also satisfied by a situation in which the urn contains seven yellow balls and one red ball.
\end{enumerate}

\subsection*{For each lottery, you must determine in advance:}
\begin{enumerate}
\item How much you are willing to stake in order to be able to participate in the subsequent \emph{real} lottery draw.
\item Which of the considered colors is the winning color for you (unless it is explicitly stated in the task that the red color is winning).
\end{enumerate}

In the first part of the experiment, you thus indicate how you would behave in the given situation. In the second part of the experiment, all online lotteries/situations are then actually realized and drawn. (They cannot be realized during data collection because the individual situations are presented to you in a random order.) Therefore, when entering your decisions into the computer, please carefully consider each situation, as you have a chance to win, but also to lose points. These points are accumulated over the four subsequent experiments, and you may win very attractive prizes. (It depends solely on you how much you stake on each lottery; a total of 14 lotteries are prepared during the experiment.)

As stated above, only 50\% of participants take part in each draw. \textbf{The higher the amount you stake, the higher the chance that you will actually participate in the draw}. On the other hand, you may lose this amount if you are selected for the draw and the color you chose is not drawn. Once the draw takes place, you no longer have the option to withdraw from the game. If the computer assigns you to the game, it deducts the number of points corresponding to the amount you were willing to \emph{pay} for participation in the given lottery. If you staked on the color that is actually drawn when the lottery is realized, you win and the computer credits you with 100 points.

\textbf{After all four experiments have been completed, your points from the individual sessions will be summed. In order to motivate you to participate as regularly as possible, we will additionally add 100 points for each participation to this sum. Thus, those who participate in all four experiments will receive, in addition to the points won or lost in the individual experiments, an extra 400 points. Moreover, they will receive a one-time \emph{thank-you reward} of 400~CZK. Those who participate in only three experiments will receive an additional 300 points plus a \emph{thank-you reward} of 200~CZK.}

\textbf{Regardless of how many sessions you participate in, you may also obtain a prize that depends on how many points you accumulate over the course of all experiments. After all four experiments have been completed, the points from the individual sessions will be summed, and prizes will be awarded to the five participants with the highest total score:}
\begin{enumerate}
\item \textbf{1st place: 5,000~CZK}
\item \textbf{2nd place: 4,000~CZK}
\item \textbf{3rd place: 3,000~CZK}
\item \textbf{4th place: 2,000~CZK}
\item \textbf{5th place: 1,000~CZK}
\end{enumerate}

\section*{Appendix C. Exact wording of experimental situations}

The experiment was based on a fixed set of predefined situations (games), each corresponding to a unique \texttt{game\_id} in the dataset.

In each session, these situations were presented to participants as tasks in an individually randomized order. The order of tasks is recorded in the dataset. The following sections reproduce the exact wording of all situations.

All graphical representations of urns correspond exactly to those shown to participants during the experiment. The graphical elements serve as visual aids accompanying the textual task descriptions and should be interpreted jointly with them.

\subsection*{C1.1 Ignorance Red (Situation No. 1)}

\begin{minipage}{0.68\linewidth}
\textbf{Content of the lottery urn}

Possible colors: \textbf{6 (black, white, red, yellow, blue, green)}.
The total number of balls is unknown.
The numbers of balls of individual colors are unknown.

\bigskip

\textbf{Question}

You win 100 CZK if the randomly drawn ball is red.
What is the maximum amount (in CZK) you are willing to bet in order to participate in the game?
\end{minipage}
\hfill
\begin{minipage}{0.28\linewidth}
\centering
\includegraphics[width=0.5\linewidth]{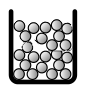}
\end{minipage}

\subsection*{C1.2 Ignorance Color (Situation No. 2)}
\begin{minipage}{0.68\linewidth}
\textbf{Content of the lottery urn}

Possible colors: \textbf{6 (black, white, red, yellow, blue, green)}
Total number of balls \textbf{is unknown}
The numbers of balls of individual colors \textbf{are unknown}

\bigskip

\textbf{Question}

Choose a color.
If the randomly drawn ball has the color you selected, you win 100 CZK.
What is the maximum amount (in CZK) you are willing to bet in order to participate in the game?

\textbf{Available color options:} red, blue, black, yellow, white, green.
\end{minipage}
\hfill
\begin{minipage}{0.28\linewidth}
\centering
\includegraphics[width=0.5\linewidth]{images/1.png}
\end{minipage}

\subsection*{C1.3 Uniform 30 Red (Situation No. 3)}

\begin{minipage}{0.68\linewidth}
\textbf{Content of the lottery urn}

Possible colors: \textbf{6 (black, white, red, yellow, blue, green)}.
The total number of balls in the urn is \textbf{30}.
The urn contains \textbf{exactly 5 balls of each color}.

\bigskip

\textbf{Question}

You win 100 CZK if the randomly drawn ball is red.
What is the maximum amount (in CZK) you are willing to bet in order to participate in the game?
\end{minipage}
\hfill
\begin{minipage}{0.28\linewidth}
\centering
\includegraphics[width=0.5\linewidth]{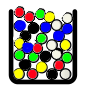}
\end{minipage}

\subsection*{C1.4 Uniform 30 Color (Situation No. 4)}

\begin{minipage}{0.68\linewidth}
\textbf{Content of the lottery urn}

Possible colors: \textbf{6 (black, white, red, yellow, blue, green)}.
The total number of balls in the urn is \textbf{30}.
The urn contains \textbf{exactly 5 balls of each color}.

\bigskip

\textbf{Question}

Choose a color.
If the randomly drawn ball has the color you selected, you win 100 CZK.
What is the maximum amount (in CZK) you are willing to bet in order to participate in the game?

\textbf{Available color options:} red, blue, black, yellow, white, green.
\end{minipage}
\hfill
\begin{minipage}{0.28\linewidth}
\centering
\includegraphics[width=0.5\linewidth]{images/3.png}
\end{minipage}

\subsection*{C2 One red in \(N\) (Situations No. 5--12)}

In later experimental sessions, the eight variants described below were presented simultaneously on a single page and were answered jointly.

\bigskip

\textbf{General description of all variants}

Possible colors: \textbf{6 (black, white, red, yellow, blue, green)}. The total number of balls is denoted by \(N\).
Exactly \textbf{one ball is red}.
The numbers of balls of the remaining colors are not specified.

\bigskip

\textbf{Question (identical for all variants)}

Choose a color.
If the randomly drawn ball has the color you selected, you win 100 CZK.
What is the maximum amount (in CZK) you are willing to bet in order to participate in the game?

\bigskip

The eight variants differed only in the total number of balls \(N\), as illustrated below.

\bigskip

\begin{center}
\setlength{\tabcolsep}{8pt}
\begin{tabular}{c c c c}

\textbf{$N=5$} &
\textbf{$N=6$} &
\textbf{$N=7$} &
\textbf{$N=8$} \\[4pt]

\includegraphics[width=0.12\linewidth]{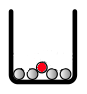} &
\includegraphics[width=0.12\linewidth]{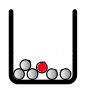} &
\includegraphics[width=0.12\linewidth]{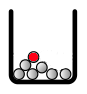} &
\includegraphics[width=0.12\linewidth]{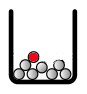} \\[12pt]

\textbf{$N=9$} &
\textbf{$N=10$} &
\textbf{$N=11$} &
\textbf{$N=12$} \\[4pt]

\includegraphics[width=0.12\linewidth]{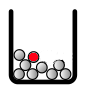} &
\includegraphics[width=0.12\linewidth]{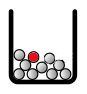} &
\includegraphics[width=0.12\linewidth]{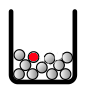} &
\includegraphics[width=0.12\linewidth]{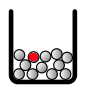} \\

\end{tabular}
\end{center}

\subsection*{C3.1 Ellsberg One Color (Situation No. 14)}

\begin{minipage}{0.68\linewidth}
\textbf{Content of the lottery urn}

Possible colors: \textbf{3 (black, red, yellow)}.
The total number of balls in the urn is \textbf{15}.
The urn contains \textbf{exactly 5 red balls}.
The remaining balls are yellow or black in an unknown proportion.

\bigskip

\textbf{Question}

Choose a color.
If the randomly drawn ball has the color you selected, you win 100 CZK.
What is the maximum amount (in CZK) you are willing to bet in order to participate in the game?

\textbf{Available color options:} red, black, yellow.
\end{minipage}
\hfill
\begin{minipage}{0.28\linewidth}
\centering
\includegraphics[width=0.5\linewidth]{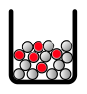}
\end{minipage}

\subsection*{C3.2 Ellsberg Two Colors (Situation No. 15)}

\begin{minipage}{0.68\linewidth}
\textbf{Content of the lottery urn}

Possible colors: \textbf{3 (black, red, yellow)}.
The total number of balls in the urn is \textbf{15}.
The urn contains \textbf{exactly 5 red balls}.
The remaining balls are yellow or black in an unknown proportion.

\bigskip

\textbf{Question}

Choose a color.
If the randomly drawn ball is yellow or has the color you selected, you win 100 CZK.
What is the maximum amount (in CZK) you are willing to bet in order to participate in the game?

\textbf{Available color options:} red, black.
\end{minipage}
\hfill
\begin{minipage}{0.28\linewidth}
\centering
\includegraphics[width=0.5\linewidth]{images/15.png}
\end{minipage}

\section*{Appendix D. Post-experimental uncertainty assessment (Experiment II only)}

The following two questions were presented at the end of each online session (Experiment II), after completion of all lottery-based decision tasks. Participants selected exactly one option. For usability reasons, the eight One-red-in-$N$ variants (Tasks 5--12) were presented jointly within a single interface block labeled ``Situation 13'' in the application. In the questionnaire responses, selection of this block is recorded as \texttt{answer = 13}, although no separate \texttt{game\_id = 13} exists in the main response table.

\subsection*{D1. Perceived greatest uncertainty}

\begin{minipage}{0.95\linewidth}
\textbf{Question}

In which situation did you perceive the \textbf{greatest uncertainty} regarding the potential win (i.e., the least information for your decision-making)?

\bigskip

\textbf{Answer options}
\begin{enumerate}
    \item Unknown number of balls in six colours, unknown colour proportions, you bet on red.
    \item Unknown number of balls in six colours, unknown colour proportions, you bet on a colour of your choice.
    \item 30 balls in six colours, 5 balls of each colour, you bet on red.
    \item 30 balls in six colours, 5 balls of each colour, you bet on a colour of your choice.
    \item \textbf{$N$} balls in six colours (\textbf{$N$} gradually takes values from 5 to 12), exactly one ball is red, you bet on a colour of your choice.
    \item 15 balls in three colours (red, black, yellow), 5 balls are red, you bet on a colour of your choice.
    \item 15 balls in three colours (red, black, yellow), 5 balls are red, you bet on a colour of your choice. You win if yellow or your chosen colour is drawn.
\end{enumerate}
\end{minipage}

\subsection*{D2. Perceived least uncertainty}

\begin{minipage}{0.95\linewidth}
\textbf{Question}

In which situation did you perceive the \textbf{least uncertainty} regarding the potential win (i.e., the most information for your decision-making)?

\bigskip

\textbf{Answer options}
\begin{enumerate}
    \item Unknown number of balls in six colours, unknown colour proportions, you bet on red.
    \item Unknown number of balls in six colours, unknown colour proportions, you bet on a colour of your choice.
    \item 30 balls in six colours, 5 balls of each colour, you bet on red.
    \item 30 balls in six colours, 5 balls of each colour, you bet on a colour of your choice.
    \item \textbf{$N$} balls in six colours (\textbf{$N$} gradually takes values from 5 to 12), exactly one ball is red, you bet on a colour of your choice.
    \item 15 balls in three colours (red, black, yellow), 5 balls are red, you bet on a colour of your choice.
    \item 15 balls in three colours (red, black, yellow), 5 balls are red, you bet on a colour of your choice. You win if yellow or your chosen colour is drawn.
\end{enumerate}
\end{minipage}

\end{document}

%% file: images/fig.tex
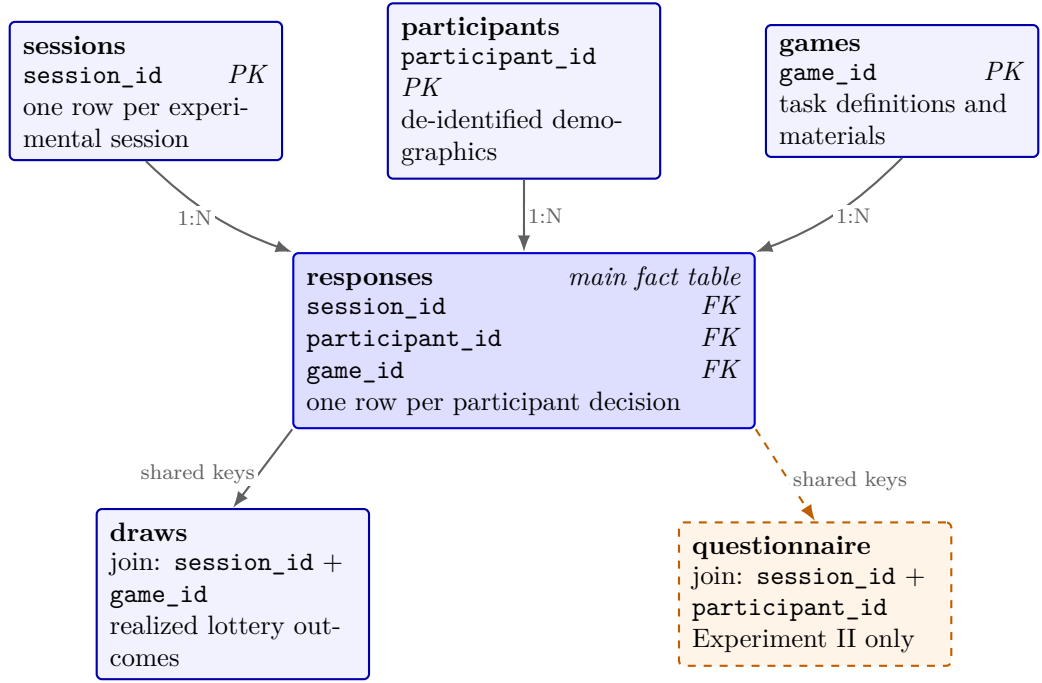
\begin{figure}[ht]
\centering
\begin{tikzpicture}[
    entity/.style={
        rectangle,
        rounded corners=2pt,
        draw=blue!65!black,
        line width=0.8pt,
        fill=blue!5,
        align=left,
        inner sep=5pt,
        text width=3.25cm,
        minimum height=1.45cm,
        font=\small
    },
    fact/.style={
        entity,
        draw=blue!80!black,
        fill=blue!13,
        text width=5.75cm,
        minimum height=1.65cm
    },
    online/.style={
        entity,
        draw=orange!75!black,
        fill=orange!9,
        dashed
    },
    link/.style={-{Latex[length=2.2mm]}, line width=0.8pt, draw=gray!75!black},
    onlineLink/.style={link, dashed, draw=orange!75!black},
    cardinality/.style={font=\scriptsize, fill=white, inner sep=1.2pt, text=gray!75!black},
    note/.style={font=\scriptsize, text=gray!70!black, align=center}
]

\node[entity] (sessions) at (-5.0,3.65) {
  \textbf{sessions}\\[-1pt]
  \texttt{session\_id} \hfill \textit{PK}\\
  one row per experimental session};

\node[entity] (participants) at (0,3.65) {
  \textbf{participants}\\[-1pt]
  \texttt{participant\_id} \hfill \textit{PK}\\
  de-identified demographics};

\node[entity] (games) at (5.0,3.65) {
  \textbf{games}\\[-1pt]
  \texttt{game\_id} \hfill \textit{PK}\\
  task definitions and materials};

\node[fact] (responses) at (0,0.35) {
  \textbf{responses} \hfill \textit{main fact table}\\[-1pt]
  \texttt{session\_id} \hfill \textit{FK}\\
  \texttt{participant\_id} \hfill \textit{FK}\\
  \texttt{game\_id} \hfill \textit{FK}\\
  one row per participant decision};

\node[entity] (draws) at (-3.85,-3.0) {
  \textbf{draws}\\[-1pt]
  join: \texttt{session\_id} +\\
  \texttt{game\_id}\\
  realized lottery outcomes};

\node[online] (questionnaire) at (3.85,-3.0) {
  \textbf{questionnaire}\\[-1pt]
  join: \texttt{session\_id} +\\
  \texttt{participant\_id}\\
  Experiment II only};

\draw[link] (sessions.south) to[bend right=12]
  node[cardinality, pos=.52, left] {1:N} (responses.north west);
\draw[link] (participants.south) --
  node[cardinality, pos=.50, right] {1:N} (responses.north);
\draw[link] (games.south) to[bend left=12]
  node[cardinality, pos=.52, right] {1:N} (responses.north east);

\draw[link] (responses.south west) --
  node[cardinality, pos=.56, left] {shared keys} (draws.north);
\draw[onlineLink] (responses.south east) --
  node[cardinality, pos=.56, right] {shared keys} (questionnaire.north);

\node[note] at (0,-4.55) {
  PK = primary key; FK = foreign key. Dashed orange elements occur only in Experiment II.};

\end{tikzpicture}

\caption{Relational structure of the released workbooks. Experiment I contains the five solid tables; Experiment II adds the dashed questionnaire table. The \texttt{responses} table links each decision to a session, participant, and game. The lower tables join to it through the displayed shared identifiers.}
\label{fig:dataset-structure}
\end{figure}